\documentclass[prl,twocolumn,amsmath,amssymb]{revtex4}

\usepackage{graphicx}
\usepackage{subfigure}
\usepackage{dcolumn}
\usepackage{array}
\usepackage{bm}
\usepackage[pass]{geometry}

\begin{document}

\title{\bf{Chern insulator at a magnetic rocksalt interface} \\[11pt] }
\author{Kevin F. Garrity and David Vanderbilt}
\affiliation{Department of Physics and Astronomy\\
Rutgers University, Piscataway, NJ 08854}

\date{\today}
\begin{abstract}

Considerable efforts have recently been devoted to the experimental
realization of a two-dimensional Chern insulator, i.e., a system
displaying a quantum anomalous Hall effect.
However, existing approaches such as those based on magnetic
doping of topological-insulator thin films have resulted in small
band gaps, restricting the effect to low temperatures.
We use first-principles
calculations
to demonstrate that an interface between thin films of the
topologically trivial ferromagnetic insulators EuO and GdN can result in a
band inversion and a non-zero Chern number.  Both materials are
stoichiometric and the interface is non-polar and lattice-matched,
which should allow this interface to be achievable experimentally.  We
show that the band structure can be tuned by layer thickness or
epitaxial strain, and can result in Chern insulators with gaps of over
0.1 eV.
\end{abstract}


\maketitle

Recently, there has been considerable interest in experimental attempts
to realize two-dimensional
Chern insulators~\cite{expt_qah, stm_magti, gddope_magti, gdn_ti},
also known as a quantum anomalous Hall insulators.  A Chern
insulator, which by definition is an insulator whose occupied bands
carry a non-zero Chern number~\cite{tknn}, is a species of
topological insulator with broken time
reversal (TR) symmetry.  Such a material displays the essential physics
of the integer quantum Hall effect, including robust dissipationless
quantized conducting edge states, but without an external magnetic
field.  These edge states could be used in electronics applications,
and Chern insulators would display strong magnetoelectric effects\cite{ti_review}.

While the requirements for a Chern insulator, namely broken TR
symmetry and spin-orbit coupling (SOC), are in principle very common,
finding a robust experimental realization of a Chern insulator has
proven very challenging.  The first successful demonstration
consisted of doping a two-dimensional slab of the $Z_2$
topological insulator (Bi,Sb)$_2$Te$_3$ with Cr atoms~\cite{expt_qah}.
The Cr dopants order ferromagnetically at low temperatures, breaking TR
symmetry and driving the system into a Chern-insulating state.
Unfortunately, the magnetic ordering temperature of this system is
only 15\,K, and $\sigma_{xy}$ becomes quantized to better than
10\% only below about 400\,mK,
greatly limiting potential applications.  Furthermore,
careful control of the Bi and Sb concentrations is needed to separate
the surface states from the bulk states, and gating is needed to adjust
the Fermi level
into the gap.  Similar proposals that focus on magnetically doping 2D
slabs of other topological insulators~\cite{magdope_bi2se3}, quantum
spin hall insulators~\cite{magdope_hgte}, or
graphene~\cite{magdope_graphene} will also likely be limited to very
low temperatures and small band gaps.

In this work, we focus on a particular realization of an alternate
strategy for designing Chern insulators.  This strategy, which
consists of directly combining magnetic insulators with materials
that have large SOC, has the advantage that
we are not limited to known topologically non-trivial materials.
As a result, we can choose
to combine materials with large band gaps and robust magnetic
orderings, potentially allowing for high-temperature operation.  We
have previously used this strategy to show that placing heavy adatoms
on the surface of a magnetic insulator frequently produces
topologically non-trivial
band structures, some of which are insulating~\cite{heavy}.  In this
work, we take the complementary
strategy of combining two trivial
magnetic insulators in such a way as to get a non-trivial band
crossing at the interface between them, resulting in a Chern
insulator.

We focus on the interface between two magnetic rock-salt materials,
EuO and GdN.  EuO is well known for being one of the few
ferromagnetically ordered semiconductors, with a Curie temperature of
70K and a band gap of 1.1\,eV~\cite{euo}.  Similarly, GdN is
ferromagnetically ordered with a Curie temperature of 58K~\cite{gdn},
and while it is a metal in the bulk~\cite{gdn2}, it is known to
become insulting in thin-film form~\cite{gdn3}.  Experimentally, the
lattice constant of EuO is 5.14\,\AA~\cite{euo_lc}
and that of GdN is 4.99\,\AA~\cite{gdn}, resulting in an
experimentally reasonable 3.0\% mismatch.  We show that
for a proper choice of strain and layer thickness, the interface
between EuO and GdN can result in a Chern-insulating state with a band
gap of over 0.1\,eV.  While the Curie temperatures of these materials
still fall below room temperature, such a system would be a significant
step in the direction of a robust room-temperature Chern insulator.

We perform first-principles density-functional theory (DFT)~\cite{hk,ks}
calculations using the VASP code~\cite{vasp, vasp2} and PAW
potentials~\cite{paw,paw2}.  We carry out structural relaxations using the PBE
generalized-gradient approximation (GGA)~\cite{pbe}, with a Hubbard U
(DFT+U) ~\cite{ldaplusU, ldaplusU_simplified} correction of 6.5 eV on
the Eu and Gd $f$ states.  Because the band alignments at the interface
are of special importance in this work, we then perform a final
band-structure calculation using a hybrid functional.  We adopt
the HSE functional~\cite{hse}, which
is known to provide a more accurate description of the band structure
of bulk EuO and GdN than DFT+U alone~\cite{euo_hybrid, gdn_hybrid, gapnote}.  
%
%
Results from our
DFT calculations are then used as input to construct
maximally-localized Wannier functions using WANNIER90
~\cite{mlwf, wannier90}.  Chern numbers and band gaps are calculated
using Wannier interpolation of the band structure; the Chern numbers
are computed by sampling the Brillouin zone (BZ) by a dense k-point
grid and summing the Berry phases around the loops formed by each
set of four adjacent k-points.

\begin{figure}
\includegraphics[width=3.5in]{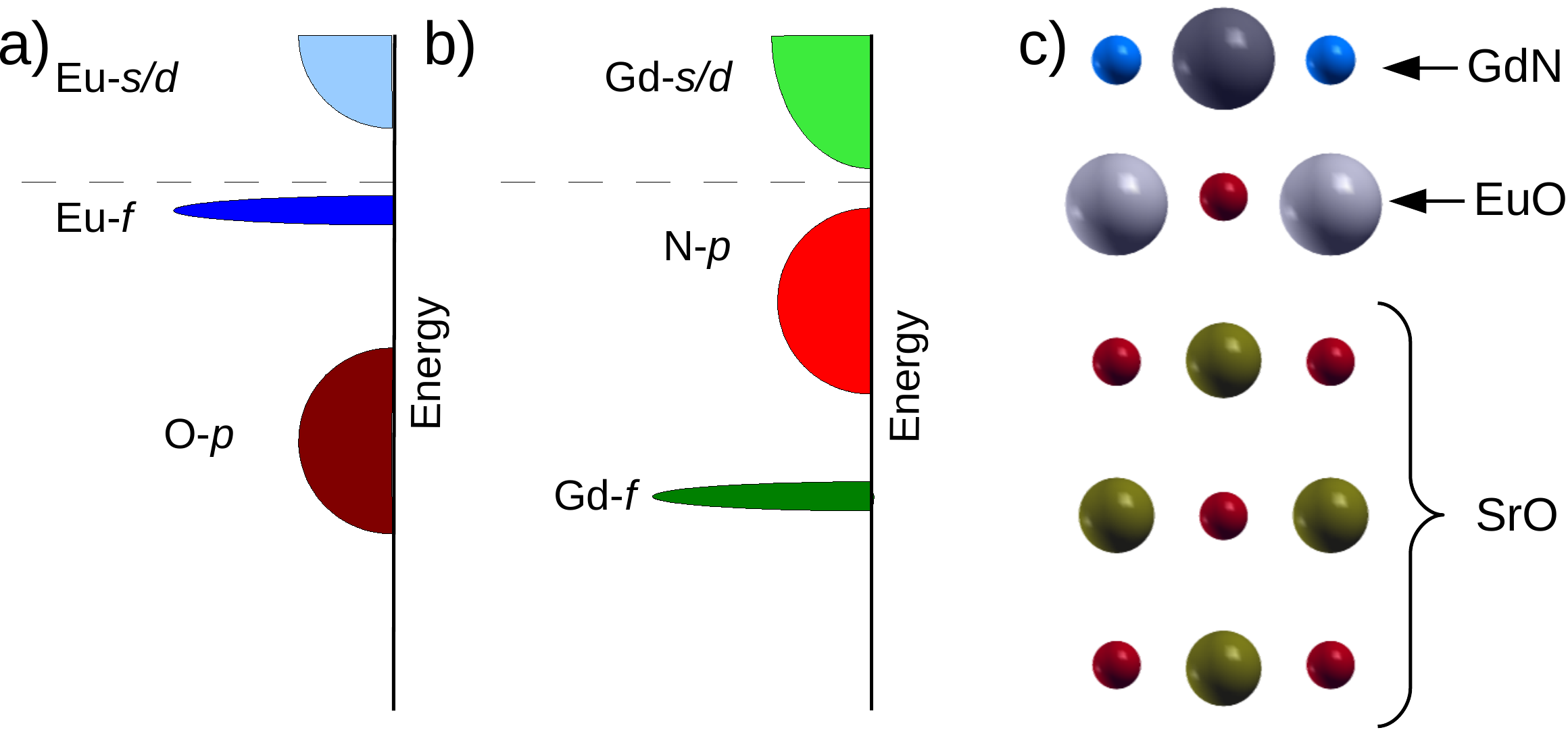}
\caption{ (a) Schematic density of states of EuO.
(b) Same for GdN.
(c) Side view of proposed interface structure, comprised of
one GdN layer atop one EuO layer on a SrO substrate.
Large atoms are cations (Gd, Eu and Sr); small atoms are
anions (N and O).}
\label{fig:schem}
\end{figure}

We begin by outlining our basic strategy for constructing a Chern
insulator.  Our goal is to find two topologically trivial materials
that, when placed together, result in a band inversion and a Chern
insulator. The (001) surface of EuO is nearly unique among binary
compounds in providing a simple insulating non-polar surface that
breaks time-reversal symmetry~\cite{spaldin}, 
making it a good starting point for our
strategy.  Both the valence and conduction bands of EuO
consist of strongly spin-polarized bands localized on the Eu atom, as shown
schematically in Fig.~\ref{fig:schem}(a).  Eu ($Z\!=\!63$) provides strong
SOC, making EuO an excellent candidate material.  In fact,
our calculations show that a single layer of EuO will become a Chern
insulator under sufficiently strong compressive epitaxial strain,
which reduces the band gap and eventually causes the conduction band
to overlap with the valence band. Unfortunately, the strain necessary to
cause this band inversion is unrealistically large ($\sim$10\%),
forcing us to look for a second material to combine with EuO to
achieve the same effect under more realistic conditions. This second
material should provide conduction bands of the correct symmetry
such that they result in an avoided crossing and a transfer of Chern
number when they overlap with the occupied Eu $f$ states.

There are many candidate materials with the rock-salt structure that
may be possible to interface epitaxially with EuO, but the
majority of them are highly ionic materials with very large band gaps,
making them unsuitable for this application.  One exception is
CdO, which has been studied previously in quantum wells and
superlattices with EuO~\cite{euo_cdo}.
We find that this combination does produce the desired
Chern-insulating state for certain values of layer thickness and
strain, but with a small band gap.  The small gap results from the
weak interaction between the Cd $s$ and Eu $f$ states near
$\Gamma$, which can be understood in the context of the
Wigner-Eckhart theorem, i.e., a first-order $\bf k\cdot p$ perturbation
($\Delta l\!=\!1$) cannot link $l\!=\!0$ and $l\!=\!3$ states,
at least without assistance from the cubic crystal field
($\Delta l\!=\!2$).
In addition, the Cd $s$ states have only a weak spin splitting
arising from exchange coupling to the Eu; this results in a
very limited phase space for non-trivial topological behavior
before the Cd $s$ state of opposite spin crosses the Fermi level
and the system becomes metallic.  Finally, CdO is
poorly lattice-matched with EuO, which would make synthesis of these
structures challenging.

Based on the example of CdO as well as general considerations of
experimental feasibility, we conclude that the ideal rock-salt material
to pair with EuO would have the following characteristics: 1) a similar
lattice constant to EuO, 2) a large SOC, 3) a small band
gap, 4) spin-polarized conduction bands with $d$-character, and 5) a
conduction band minimum at $\Gamma$ in the surface BZ.  The
only material we know of that meets all of these requirements is GdN,
a ferromagnetic spin-polarized semi-metal in bulk which becomes
insulating in thin-film form.

\begin{figure}
\includegraphics[width=3.5in]{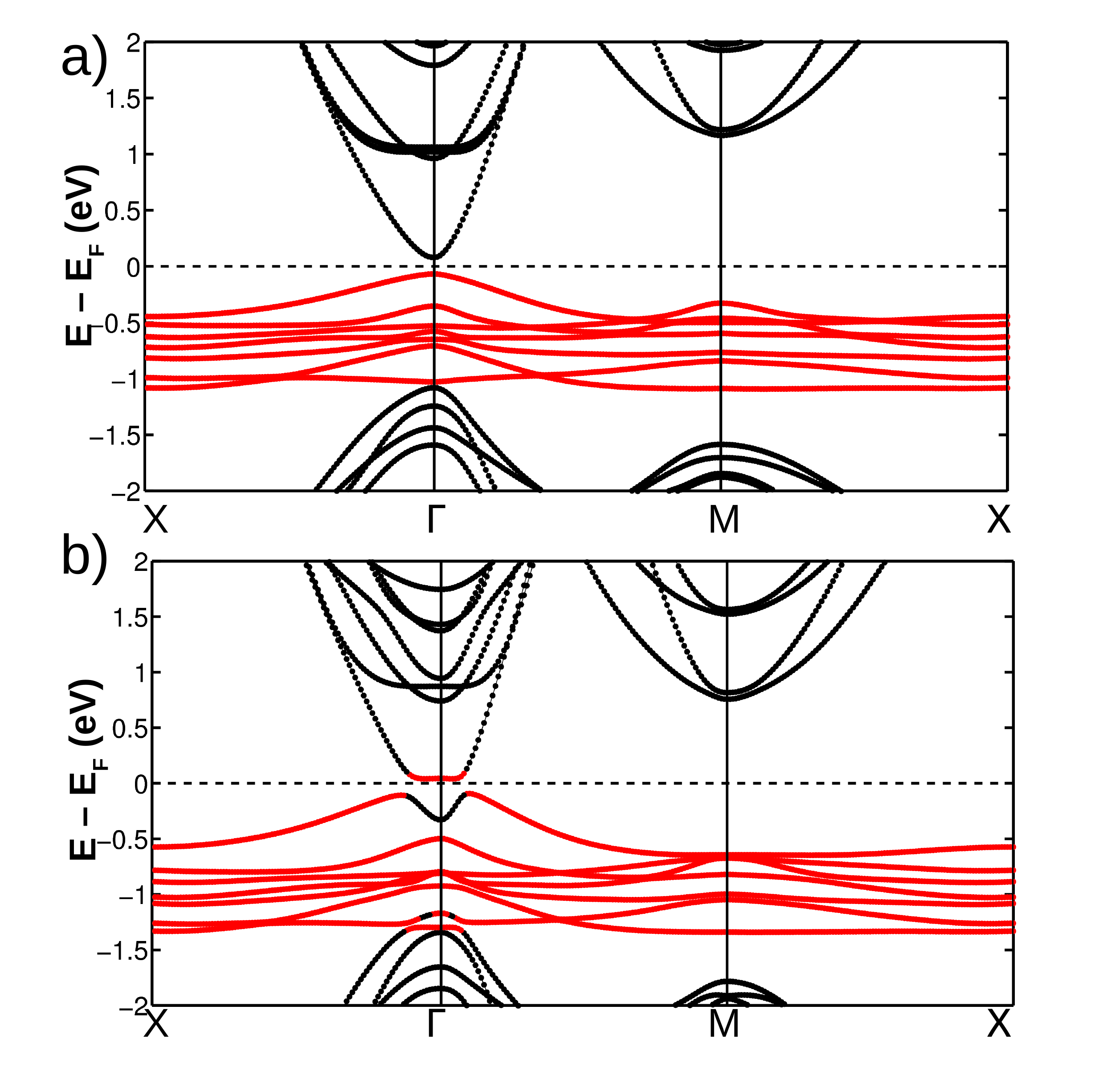}
\caption{Band structures of one GdN layer atop one EuO layer on an SrO
substrate at
(a) 3.62\,\AA\ ($C\!=\!0$), and
(b) 3.44\,\AA\ ($C\!=\!-1$). Majority Eu $f$ states are plotted
in red; others are in black.}
\label{fig:bs1}
\end{figure}

In Fig.~\ref{fig:schem}(a-b) we present a schematic band-alignment
diagram for
GdN and EuO.  We seek to engineer a band inversion between the Eu $f$
state at the top of the valence band of EuO and the Gd $d_{x^2-y^2}$
orbital at the bottom of the conduction band of GdN by creating an
interface between these two materials.  In order to obtain the largest
possible coupling between the two materials, we begin with single
layers of EuO and GdN stacked on an SrO substrate, as shown in
Fig.~\ref{fig:schem}(c).  We show the resulting surface band structure
in Fig.~\ref{fig:bs1}, with all spins aligned ferromagnetically
perpendicular to the surface, for two different values of in-plane
lattice constant.  At large lattice constants this interface is a
trivial insulator, as shown in Fig.~\ref{fig:bs1}(a), but reducing
the lattice constant to 3.53\,\AA\ causes
the band gap to close.  At this critical strain, the Gd $d_{x^2-y^2}$
conduction-band minimum crosses the Eu $f_{x^3-i y^3}$
valence-band maximum at $\Gamma$.  Further reduction of the lattice
constant, as shown in Fig.~\ref{fig:bs1}(b), results in an avoided
crossing and a transfer of Chern number from the valence band to the
conduction band, leaving the occupied bands with a Chern number of $-1$.

By varying the strain in the system, the band gap can be increased
above 0.1 eV, as shown in the first three rows of Table~\ref{tab:chern}.
This very robust gap can be traced to several
factors.  First, the atomically thin layers of GdN and EuO confine
the relevant states at the interface, resulting in a large overlap.
Second, as discussed above, the $f$ and $d$ character of the
bands allows for first-order
coupling between them.  Finally, the large separation of the
conduction-band minimum and valence-band maximum from other bands in
the system allows for a large band inversion and a strong avoided
crossing.  This isolation of the relevant bands also allows for a
relatively wide range of strains that can result in a Chern-insulating
state, which should make our predictions more robust and easier to
achieve experimentally.

\begin{table} 
\begin{center} 
\begin{ruledtabular} 
\begin{tabular}{ccddd}
$N_{\rm EuO}$ & $N_{\rm GdN}$ & \multicolumn{1}{c}{$a$ (\AA)} &
\multicolumn{1}{c}{$C$} & \multicolumn{1}{c}{$E_{\rm g}$ (meV)} \\
\hline
1 & 1 & 3.62 & 0 &  111 \\
1 & 1 & 3.53 & -1 &   3 \\
1 & 1 & 3.48 & -1 & 130 \\
2 & 1 & 3.62 & -1 &  80 \\
2 & 1 & 3.53 & -1 & 123 \\
1 & 2 & 3.48 & -1 & 71 \\
2 & 2 & 3.62 & -1 & 62 \\
\end{tabular} 
\end{ruledtabular} 
\caption{Computed Chern number $C$ and band gap $E_{\rm g}$ for
$N_{\rm GdN}$ layers of GdN atop $N_{\rm EuO}$ layers of EuO on
an SrO substrate of the specified lattice constant $a$.}
\label{tab:chern} 
\end{center} 
\end{table}

The value of the Chern number in this system can be understood by examining the
symmetries of the bands that take part in the avoided
crossing~\cite{chern_band_cross}.  At the $\Gamma$ point
of the BZ, the bands all belong to one of four
non-degenerate irreducible representations that can be labeled by the
eigenvalues of the four-fold rotation operator. In the case of a
single band crossing where the eigenvalues of the bands differ
only by a factor of $\pm i$,
the exchange of Chern numbers is uniquely
determined~\cite{chern_numbers}.  In the case of
GdN on EuO, where the conduction-band minimum has the symmetry of a
$d_{x^2-y^2}$ orbital, these symmetry considerations imply that $C\!=\!-1$,
consistent with our direct numerical
calculations.  In the case of CdO, the conduction band minimum has
$s$-character, and $C\!=\!1$.

In order to gain some insight into the behavior of this system, we
build a simple two-band single-spin tight-binding model,
in the spirit of the Haldane model~\cite{haldane}, but here designed
to capture the key behavior of the valence-band maximum and
conduction-band minimum of our system.
The model consists of a square lattice with a
single $d_{x^2-y^2}$ orbital at $(0,0)$ and a single $f_{x^3-i y^3}$
orbital at $(\frac{1}{2},\frac{1}{2})$, labeled by $a\,=\,\{1,2\}$.
These orbitals have on-site energies
$\pm \Delta$, and strong hoppings ($t_1\!<\!0, t_2\!>\!0$)
to nearest neighbors of the same sublattice.  The sublattices are
coupled by a weaker complex interaction term,
$\lambda_{ij} = u \, \hat{e}_{ij}\cdot(\hat{x} - i \hat{y})$,
where $u$ is the magnitude of the coupling and
%
%
$\hat{e}_{ij}$
is the direction of the hopping.  This results in the Hamiltonian
\begin{eqnarray}\label{nonan}
H = \sum_{i,a} (-1)^{a+1} \Delta c_{ia}^{\dag} c_{ia} + \sum_{<ij>a}
t_a c_{ia}^{\dag} c_{ja} \nonumber \\
 + \sum_{<ij>,a,b} \lambda_{ij}
c_{ia}^{\dag} c_{jb} + H.c.  
\end{eqnarray} 
The sums over $i$ and $j$ are over unit cells, and the sums over $a$
and $b$ are over the two orbitals.  By reducing $\Delta$, which mimics
the impact of strain on the GdN/EuO system, the model can be tuned
from a trivial to a Chern-insulating state, with the two phases
separated by a band touching at $\Gamma$ at the critical value of
$\Delta_c\!=\!-2 t_1\!+\!2 t_2$.  For $\Delta$ slightly below $\Delta_c$,
the band gap of the Chern insulating phase increases linearly with
$\Delta_c\!-\!\Delta$, but for larger band inversions, the band gap of
the Chern-insulating state begins to saturate, growing like
$(\Delta_c\!-\!\Delta)^{\frac{1}{2}}$ as the avoided crossings move away from
$\Gamma$.  This behavior is
consistent with our first-principles results showing increasing gaps
for larger band inversions, as long as the two orbitals near the Fermi
level remain separated from other bands in the system.

\begin{figure}
\includegraphics[width=3.5in]{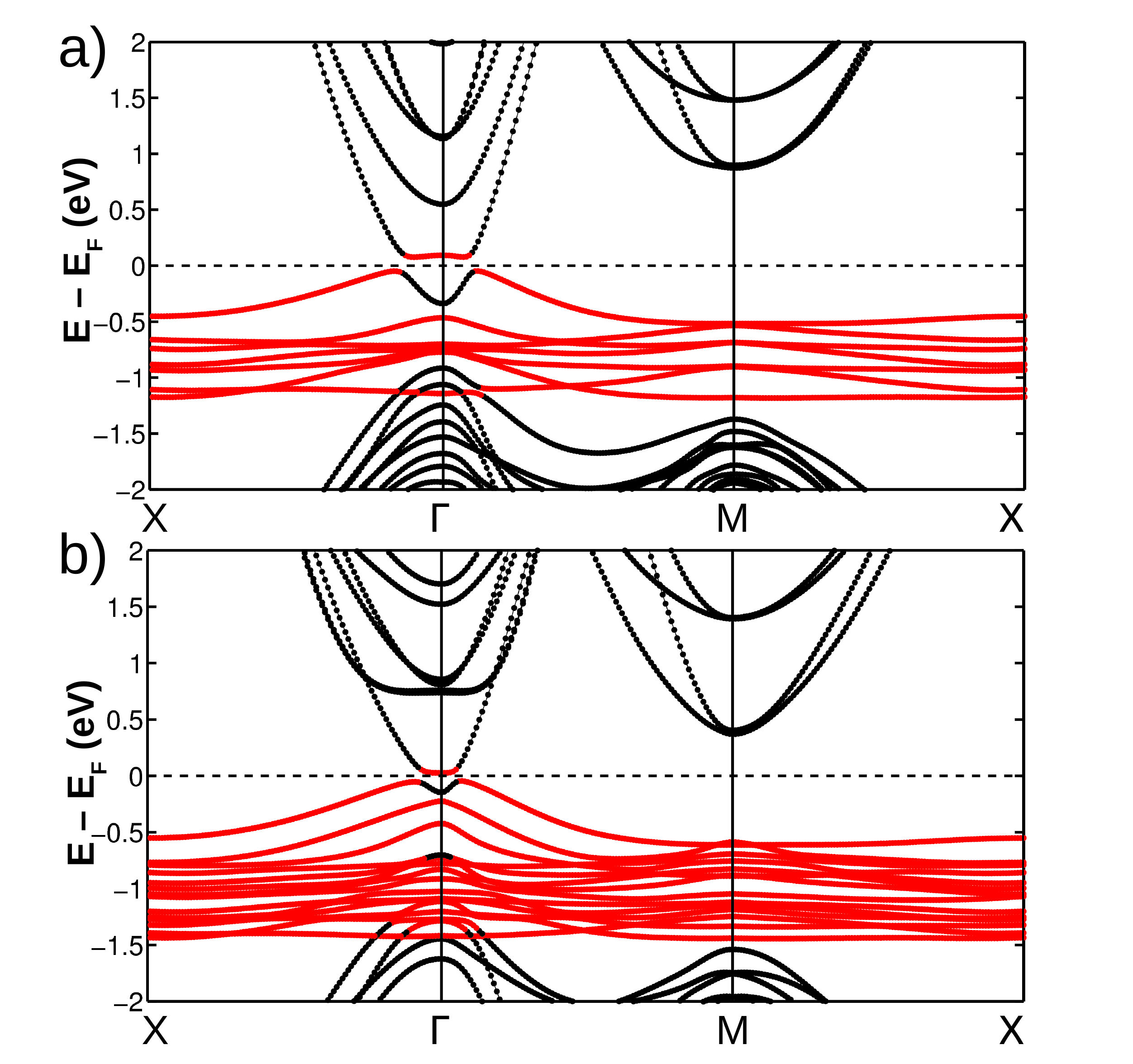}
\caption{Band structures of
(a) two GdN layers atop one EuO layer on SrO at 3.53\,\AA, and
(b) one GdN layer atop two EuO layers on SrO at 3.48\,\AA.
Both have $C\!=\!-1$.  Colors as in Fig.~\ref{fig:bs1}.}
\label{fig:bs2}
\end{figure}

Returning to the first-principles results, we next consider thicker
layers of both GdN and EuO.  These thicker layers are likely to be
easier to grow and measure experimentally, as well as more likely
to have bulk-like magnetic ordering temperatures.  First, we consider
an additional GdN layer, as shown in Fig.~\ref{fig:bs2}(a).  Consistent
with our expectations, an additional layer of GdN results in a lower
conduction-band minimum at a given lattice constant,
consistent with the fact
that thicker slabs of GdN should approach bulk-like semi-metallic
behavior.  This behavior is convenient because it allows the band
structure to be tuned not only by strain, but also by varying the
GdN layer thickness.
We expect that as the GdN thickness increases, the
relevant conduction-band state will become more delocalized, leading
to less interaction with EuO and a smaller band gap. However, this
effect appears to be fairly weak when going from one to two layers,
as shown in Table~\ref{tab:chern}.

We also explore the band structure of a single GdN layer on top of two
EuO layers, as shown in Fig.~\ref{fig:bs2}(b).  This configuration also
results in a Chern insulator for appropriate values of strain, but
the additional Eu $f$ levels from the second EuO layer limit
the total band gap.  This occurs because of the relatively weak
splitting between the $f$ orbitals on the two Eu atoms; for large
band inversions, the $d_{x^2-y^2}$ band from the Gd crosses several of
the Eu-$f$ states, closing the band gap.  However, as shown in
Table~\ref{tab:chern}, the splitting between different Eu $f$ levels is
still large enough to allow significant band gaps, even for
thicker EuO layers.

In all of the above calculations, we have assumed that the spins in
both EuO and GdN align ferromagnetically, as they do in the bulk.
As a preliminary test of this hypothesis, we consider an
antiferromagnetic arrangement with a single spin-up GdN layer
atop a single spin-down EuO layer on SrO at 3.53\,\AA, and we find that
the antiferromagnetic configuration is 0.5\,meV higher in energy
than the ferromagnetic configuration
considered previously.  This energy difference is consistent, in
sign and magnitude, with the calculated exchange parameters in bulk GdN and
EuO~\cite{euo_hybrid, gdn_hybrid}.  Further calculations would be
needed to determine all of the relevant magnetic exchange
parameters, as well as the effects of reduced dimensionality on the
magnetic ordering temperature.  In addition, we note that we have
assumed that the spins align in the $z$ direction; experimentally,
this might require the application of a small external field.
Nevertheless, with these qualifications, we believe the needed
magnetic structures should be attainable experimentally.

To summarize, our proposal results from a design strategy in which known
topologically-trivial materials with strong spin-orbit coupling
are combined in such a way as to to engineer a band inversion,
resulting in robust, topologically non-trivial behavior.
We have shown that the (001) interface between GdN and EuO is an
excellent candidate system for achieving a robust Chern-insulating
state at temperatures of up to 70K and with band gaps of over 0.1\,eV.
This non-polar lattice-matched interface consists of known
stoichiometric magnetic insulators, which should make it achievable
experimentally.

\vspace{0.3cm}
\noindent{\bf Acknowledgments}
\vspace{0.3cm}

This work was supported by NSF Grant DMR-10-05838.


\begin{thebibliography}{35}
\bibitem[{gap()}]{gapnote}
\bibinfo{note}{Unlike PBE0, HSE underestimates the EuO band gap, but in testing
  we found that calculations with PBE0 and HSE give very similar results near
  the Fermi level for GdN/EuO interfaces.}
\end{thebibliography}

\end{document}